\documentstyle [12pt] {article}
\begin{document}

\pagestyle{empty}
\rightline{\vbox{
	\halign{&#\hfil\cr
	&NUHEP-TH-93-5\cr
	&April 1993\cr}}}
\vskip 1in
{\Large\bf
\centerline{Tau Polarimetry with Inclusive Decays}}
\vskip .5in
\normalsize
\centerline{Eric Braaten}
\centerline{\sl Department of Physics and Astronomy, Northwestern University,
    Evanston, IL 60208}
\vskip 1.5in
\begin{abstract}
The spin asymmetry parameter $A_\tau$ characterizing the angular distribution
of the total hadron momentum in the decay of a polarized tau
can be calculated rigorously using perturbative QCD and the operator product
expansion.  Perturbative QCD corrections to the free quark result $A_\tau =
1/3$
can be expressed as a power series in $\alpha_s(M_\tau)$ and nonperturbative
QCD corrections can be expanded systematically in powers of $1/M_\tau^2$.
The QCD prediction is $A_\tau = 0.41 \pm 0.02$.
In the decay of a high energy tau into hadrons, the value of the
hadronic energy distribution $dR_\tau/dz$ evaluated at the maximum hadronic
energy fraction $z = 1$ can also be calculated rigorously from QCD.

\end{abstract}
\vfill \eject \pagestyle{plain} \setcounter{page}{1}

The spin-dependence of processes involving elementary particles
contains a wealth of information about their fundamental interactions.
Unfortunately this information is not easily accessible to experiment.
It requires either the use of polarized beams and targets, or the
measurement of the polarization of final state particles.
The tau lepton is one of the few elementary particles whose
spin can be effectively analyzed by its decays.
It is well known that several of the exclusive decay modes of the tau
can be used to analyze its spin \cite{tsai,hmz}.
In this Letter, I point out that inclusive decays into hadrons
can also be used for this purpose.  The polarization of a sample of taus
that decay into hadrons can be determined from the angular distribution
of the total momentum of the hadrons.
The asymmetry parameter that characterizes the angular distribution
can be computed systematically using perturbative QCD and the operator
product expansion.  In a sample of high energy taus that decay into hadrons,
the endpoint value of the hadronic energy distribution
can also be computed rigorously from QCD, and can be used to measure the
average helicity of the taus.  Measurements of these spin-dependent
observables could provide dramatic confirmation of the applicability of
perturbative QCD to the inclusive hadronic decays of the tau.

It is convenient to normalize the inclusive decay
rate of the tau lepton into a neutrino plus hadrons
to the electronic decay rate by defining the ratio
\begin{equation} {
R_\tau \;=\; {\Gamma(\tau^- \rightarrow \nu_\tau + {\rm hadrons})
	\over \Gamma(\tau^- \rightarrow \nu_\tau e^- {\bar \nu}_e)} \;.
} \label{Rdef} \end{equation}
For a sample of taus with polarization $P$, the angle $\theta$
between the total momentum of the hadrons in the tau rest frame
and the spin quantization axis
has a distribution proportional to $1 + A_\tau P \cos \theta$,
where $A_\tau$ is an asymmetry parameter.
This angular distribution can be used to separate
$R_\tau$ into ``forward'' and ``backward'' components $R_F$ and $R_B$:
\begin{equation} {
{d R_\tau \over d \cos \theta}
	\;=\; R_F {1 + P \cos \theta \over 2}
	\;+\; R_B {1 - P \cos \theta \over 2} \;.
} \label{RFB} \end{equation}
The asymmetry parameter $A_\tau$ is then
\begin{equation} {
A_\tau \;=\; {R_F - R_B \over R_F + R_B} \;.
} \label{Adef} \end{equation}

A naive estimate of the asymmetry parameter can be obtained
by considering the decay of the tau into hadrons at the quark level,
where it proceeds through the processes
$\tau^- \rightarrow \nu_\tau d {\bar u}$ and
$\tau^- \rightarrow \nu_\tau s {\bar u}$.
The momentum of the $d {\bar u}$ and $s {\bar u}$ pairs can be identified
with the total momentum of the hadrons.  Ignoring the QCD interactions
that bind the quarks into color singlet hadrons, the angular distribution
of the total hadron momentum is
\begin{equation} {
{d R_\tau \over d \cos \theta}
\;\approx\; {3 \over 2} \left( |V_{ud}|^2 \;+\; |V_{us}|^2 \right)
\left( 1 \;+\; {1 \over 3} P \cos \theta \right) \;.
} \label{dRnaive} \end{equation}
The squares of the Kobayashi-Maskawa
matrix elements add up to 1 to high accuracy, so they will be omitted below.
The naive estimates for the ratio (\ref{Rdef}) and the asymmetry parameter
(\ref{Adef}) are $R_\tau = 3$ and $A_\tau = 1/3$.

In the case of $R_\tau$, the QCD corrections to the naive result
can be computed systematically using perturbative QCD and the
operator product expansion \cite{st,eba,np}.  The  perturbative corrections
can be expanded as a power series in $\alpha_s(M_\tau)$ \cite{ebb} and the
nonperturbative corrections can be organized systematically into
an expansion in powers of $1/M_\tau^2$.  A thorough analysis of the
QCD and electroweak corrections to the ratio $R_\tau$
has recently been carried out \cite{bnp}.
The methods that were used to calculate the ratio $R_\tau$ can also be used
for a rigorous calculation of the asymmetry parameter $A_\tau$.
The starting point is an expression for the angular distribution
of the total hadron momentum as an integral
over the invariant mass $s$ of the hadrons:
\begin{eqnarray}
{d R_\tau \over d \cos \theta}
\;=\; 6 \pi \int_0^{M_\tau^2} {ds \over M_\tau^2}
	\left( 1 - {s \over M_\tau^2} \right)^2
\Bigg( {\rm Im} \Pi^{(1)}(s + i \epsilon)
	\left( 1 + P \cos \theta
		\;+\; 2 {s \over M_\tau^2} (1 - P \cos \theta) \right)
\nonumber \\
\;+\; {\rm Im} \Pi^{(0)} (s + i \epsilon) ( 1 + P \cos \theta) \Bigg) \;,
\label{Rint} \end{eqnarray}
where $\Pi^{(J)}(s), J = 0,1$ are the transverse and
longitudinal correlators
for the quark current that couples to the virtual $W$ boson.
The notation is the same as in Ref. \cite{bnp}.
The correlators $\Pi^{(J)}(s)$ are analytic functions of $s$ except along
the positive real $s$-axis.  This allows the integral in (\ref{Rint})
to be expressed as a contour integral in the complex $s$-plane.  The contour
can be deformed so that it runs counterclockwise around the circle
$|s| = M_\tau^2$.  The resulting expressions for the forward and backward
components of $R_\tau$ defined in (\ref{RFB}) are
\begin{eqnarray}
R_F &=& {- 12 \pi^2 \over 2 \pi i}
\int_{|s|=M_\tau^2} {ds \over M_\tau^2} \left( 1 - {s \over M_\tau^2} \right)^2
	\Bigg( \Pi^{(1)}(s)  \;+\; \Pi^{(0)}(s) \Bigg) \;,
\label{RF} \\
R_B &=& {- 12 \pi^2 \over 2 \pi i}
\int_{|s|=M_\tau^2} {ds \over M_\tau^2} \left( 1 - {s \over M_\tau^2} \right)^2
	\Bigg( 2 {s \over M_\tau^2} \Pi^{(1)}(s) \Bigg) \;.
\label{RB} \end{eqnarray}
The contour integral expressions (\ref{RF}) and (\ref{RB}) reveal that the
polarization asymmetry $A_\tau$, like the ratio $R_\tau$, is completely
determined by correlation functions at the distance scale $1/M_\tau$.
This implies that the nonperturbative long distance effects of QCD
can be expressed in terms of matrix elements of local operators.
These matrix elements appear when
the operator product expansion is used to expand the correlators
$\Pi^{(J)}(s)$ in (\ref{RF}) and (\ref{RB}) in powers of $1/s$.
Evaluating the contour integrals, the QCD corrections to the naive predictions
$R_F = 2$ and $R_B = 1$  are obtained as systematic expansions
in powers of $1/M_\tau^2$.
There is also an important electroweak correction consisting of
a multiplicative short distance factor $S_{EW} = 1.019$ \cite{ms}.
The resulting expressions for the forward and backward components of
$R_\tau$ have the form
\begin{equation} {
R_F \;=\;  2 S_{EW} \;
\left(1 \;+\; \delta^{(0)}_F \;+\; \delta^{(2)}_F \;+\; \delta^{(4)}_F
	\;+\; \delta^{(6)}_F \;+\; ... \right) \;,
} \label{RFdel} \end{equation}
\begin{equation} {
R_B \;=\;  S_{EW} \;
\left(1 \;+\; \delta^{(0)}_B \;+\; \delta^{(2)}_B \;+\; \delta^{(4)}_B
	\;+\; \delta^{(6)}_B \;+\; ... \right) \;,
} \label{RBdel} \end{equation}
where the fractional corrections $\delta^{(n)}_F$ and $\delta^{(n)}_B$ are
proportional to $1/M_\tau^n$ with coefficients
that depend logarithmically on $M_\tau$.  For $R_\tau = R_F + R_B$,
the fractional corrections to the free quark value $3 S_{EW}$ are
$(2 \delta^{(n)}_F + \delta^{(n)}_B)/3$.

The fractional corrections $\delta^{(n)}_F$ and $\delta^{(n)}_B$
can be calculated straightforwardly
using the operator product expansions for the correlators $\Pi^{(J)}(s)$
that are collected in Ref. \cite{bnp}.  The dimension-0 corrections,
which represent the purely perturbative effects from the interactions
of massless quarks and gluons, are
\begin{eqnarray}
\delta^{(0)}_F &=&  {\alpha_s \over \pi}
\;+\; 5.765 \left( {\alpha_s \over \pi} \right)^2
\;+\; 34.48 \left( {\alpha_s \over \pi} \right)^3
\;+\; (d_4 + 165.1) \left( {\alpha_s \over \pi} \right)^4 \;,
\label{delF0} \\
\delta^{(0)}_B &=&  {\alpha_s \over \pi}
\;+\; 4.077 \left( {\alpha_s \over \pi} \right)^2
\;+\; 10.13 \left( {\alpha_s \over \pi} \right)^3
\;+\; (d_4 - 96.1) \left( {\alpha_s \over \pi} \right)^4 \;,
\label{delB0} \end{eqnarray}
where $\alpha_s = \alpha_s(M_\tau)$ is the running coupling constant
of QCD in the $\overline{MS}$ scheme evaluated at the scale $M_\tau$.
The coefficient $d_4$ in the $\alpha_s^4$ term in (\ref{delF0})
and (\ref{delB0}) is the fourth coefficient in the perturbative
expansion of $-2 \pi^2 s(d/ds)\Pi^{(1)}(s)$ in powers of $\alpha_s/\pi$
and has not been calculated.
The previous coefficients are 1, 1, 1.64, and 6.37 \cite{gkl}.
We assign a very conservative error to this unknown coefficient:
$d_4 = 0 \pm 100$.  The corresponding coefficient in the fractional
correction to $R_\tau$ is then $78.0 \pm 100$.
The dimension-2 corrections are perturbative corrections due to the
running quark masses.  The only correction that is numerically significant
comes from the strange quark mass $m_s = m_s(M_\tau)$:
\begin{eqnarray}
\delta^{(2)}_F &=&  - 9 \; \sin^2 \theta_C
	\; \left( 1 + {16 \over 3} {\alpha_s \over \pi} \right)
	\; {m_s^2 \over M_\tau^2} \;,
\label{delF2} \\
\delta^{(2)}_B &=&  - 6 \; \sin^2 \theta_C
	\; \left( 1 + {16 \over 3} {\alpha_s \over \pi} \right)
	\; {m_s^2 \over M_\tau^2} \;,
\label{delB2} \end{eqnarray}
where $\theta_C$ is the Cabbibo mixing angle:  $\sin^2 \theta_c = 0.049$.
For the running strange quark mass
in the $\overline{MS}$ scheme evaluated at the scale $M_\tau$,
we take the value $m_s(M_\tau) = (0.17 \pm 0.02) \; {\rm GeV}$.
The first nonperturbative corrections appear at dimension 4 in the form
of scale invariant matrix elements called the gluon condensate and quark
condensates:
\begin{eqnarray}
\delta^{(4)}_F &=& 2 \pi^2
	\left( 1 - {11 \over 18} {\alpha_s \over \pi} \right)
	{<\!\!(\alpha_s/\pi) GG\!\!> \over M_\tau^4}
\;+\; 48 \pi^2 {<\!\!m{\bar \psi} \psi\!\!> \over M_\tau^4}
\;-\; {72 \over 7} \sin^2 \theta_C
	{\pi \over \alpha_s} \; {m_s^4 \over M_\tau^4} \;,
\label{delF4} \\
\delta^{(4)}_B &=& -4 \pi^2
	\left( 1 - {11 \over 18} {\alpha_s \over \pi} \right)
	{<\!\!(\alpha_s/\pi) GG\!\!> \over M_\tau^4} \;.
\label{delB4} \end{eqnarray}
Note that the contribution of the gluon condensate $<\!\!(\alpha_s/\pi)
GG\!\!>$
cancels to order $\alpha_s$
in the fractional correction $(2 \delta^{(4)}_F + \delta^{(4)}_B)/3$
to the ratio $R_\tau$.  This results in a suppression of the gluon
condensate contribution to $R_\tau$ by two orders of magnitude.
We take the value of the gluon condensate to be
$<\!\!(\alpha_s/\pi) GG\!\!> = (2 \pm 1) \times 10^{-2} \; {\rm GeV}^4$
\cite{sn}.
The matrix element $<\!\!m{\bar \psi}\psi\!\!>$ in (\ref{delF4})
is a weighted average of the
quark condensates:
\begin{equation} {
<\!\!m{\bar \psi}\psi\!\!> \;=\;
{<\!\!m_u{\bar u}u\!\!> + \cos^2 \theta_C \; <\!\!m_d{\bar d}d\!\!>
	+ \sin^2 \theta_C \; <\!\!m_s{\bar s}s\!\!> \over 2} \;.
} \label{m2} \end{equation}
Its value is $<\!\!m{\bar \psi}\psi\!\!> = (-8 \pm 1) \times 10^{-5} \;
{\rm GeV}^4$.
The inverse power of $\alpha_s(M_\tau)$ multiplying the $m_s(M_\tau)^4$ term
in (\ref{delF4}) was first understood by Broadhurst and Generalis \cite{bg}.
At dimension 6, there are too many unknown matrix elements
for a completely systematic treatment.  Within the vacuum saturation
approximation, these corrections are
\begin{eqnarray}
\delta^{(6)}_F &=& {256 \pi^3 \over 27}
{\rho \alpha_s\!<\!\!{\bar \psi} \psi\!\!>^2 \over M_\tau^6} \;,
\label{delF6} \\
\delta^{(6)}_B &=& - {2048 \pi^3 \over 27}
{\rho \alpha_s\!<\!\!{\bar \psi} \psi\!\!>^2 \over M_\tau^6} \;,
\label{delB6} \end{eqnarray}
with $\rho = 1$.  It has been found empirically that this approximation
underestimates the size of the dimension-6 correction, and it is better to
treat
$\rho \alpha_s\!<\!\!{\bar \psi} \psi\!\!>^2$ as an effective
scale-invariant operator
of dimension 6, independent of $<\!\!{\bar \psi} \psi\!\!>$.
The best estimate for this parameter is
$\rho \alpha_s\!<\!\!{\bar \psi}\psi\!\!>^2
= (4 \pm 2) \times 10^{-4} \; {\rm GeV}^6$.
The dimension-8 and higher corrections are assumed to be completely negligible.

Inserting the fractional corrections given above into
(\ref{RFdel}) and (\ref{RBdel}), we obtain
predictions for the ratio $R_\tau = R_F + R_B$
and the asymmetry parameter $A_\tau$ defined in (\ref{Adef})
as a function of $\alpha_s(M_\tau)$ and the five parameters
$d_4$, $m_s(M_\tau)$, $<\!\!(\alpha_s/\pi) GG\!\!>$,
$<\!\!m {\bar \psi}\psi\!\!>$,
and $\rho \alpha_s\!<\!\!{\bar \psi}\psi\!\!>^2$.
Alternatively, given a value for $R_\tau$, we can predict
both $\alpha_s(M_\tau)$ and $A_\tau$.  The predictions are shown in Table 1.
The uncertainty in $\alpha_s(M_\tau)$
is dominated by the assumed error of $\pm 100$ in the coefficient $d_4$.
For $R_\tau = 3.60$, the uncertainty in $\alpha_s(M_\tau)$
is $3.6 \%$.  After $d_4$, the next largest errors are $1.1 \%$ from
$\rho \alpha_s\!<\!\!{\bar \psi}\psi\!\!>^2$ and $0.4 \%$ from $m_s(M_\tau)$.
The uncertainty in $A_\tau$ is dominated by the
gluon condensate, and is $5.4 \%$ for $R_\tau = 3.60$.
The next largest errors are $1.8 \%$ from
$\rho \alpha_s\!<\!\!{\bar \psi}\psi\!\!>^2$ and $0.7 \%$ from $d_4$.

There are two independent ways of measuring the ratio $R_\tau$ experimentally.
Using the universality of electron and muon couplings,
it can be expressed in terms of the electronic branching fraction
$B_e$ of the tau: $R_\tau = (1/B_e) - 1.973$.
Alternatively, using the universality of electron and tau couplings as well,
it can be expressed in terms of the masses and lifetimes of the
$\mu$ and $\tau$: $R_\tau = (\tau_\mu/\tau_\tau) (M_\mu/M_\tau)^5 - 1.973$.
The present world average for the electronic
branching fraction  is $B_e = (17.78 \pm 0.15) \%$ \cite{wjm},
and it gives the ratio $R_\tau = 3.651 \pm 0.047$.
The present world average for the tau lifetime is
$\tau_\tau = (2.96 \pm 0.03) \times 10^{-13} \; {\rm s}$ \cite{wjm}.
Combined with the recent precise measurement of the tau mass \cite{bs},
it gives the ratio $R_\tau = 3.545 \pm 0.056$.
Forming the weighted average of the two independent determinations of $R_\tau$,
we get $R_\tau = 3.607 \pm 0.036$.  From Table 1, this determines
the running coupling constant at the scale $M_\tau$ to be
$\alpha_s(M_\tau) = 0.319 \pm 0.017$.
We have added in quadrature the error from Table 1 and the error due
to the experimental uncertainty in $R_\tau$.  Evolving the running
coupling constant up to the $Z^0$ mass, it becomes
$\alpha_s(M_Z) = 0.1176 \pm 0.0021$.
The QCD prediction for the asymmetry parameter is
$A_\tau = 0.413 \pm 0.022$.  A measurement of $A_\tau$
consistent with this prediction would provide dramatic confirmation
of the accuracy of the QCD calculation of the ratio $R_\tau$.

The angular distribution $dR_\tau/d\cos\theta$ that defines the
asymmetry parameter $A_\tau$ is
most easily measured in low energy experiments.  In high energy experiments,
such as the decay of the $Z^0$ into taus, the quantity that is most easily
measured is the distribution $dR_\tau/dz$ of the hadron energy fraction
$z = E_H/E_\tau$, where $E_H$ is the total energy of the hadrons in the
rest frame of the $Z^0$ and $E_\tau$ is the energy of the decaying tau.
If the spin quantization axis is chosen to lie along the direction of the
tau momentum, then the variables
$z$ and $\cos \theta$ are related by
\begin{equation}{
\cos \theta \;=\; {(2z-1)M_\tau^2 - s \over M_\tau^2 - s}
} \label{zcos} \end{equation}
for a hadronic state with invariant mass $s$ \cite{hmz}.
Changing variables in (\ref{Rint}) from $\cos \theta$ to $z$,
we obtain an expression for the hadronic energy
distribution $D_P(z) = dR_\tau/dz$ for a tau
with polarization $P$ in the helicity
basis. At the endpoint $z=1$, it reduces to
\begin{eqnarray}
D_P(z=1) \;=\; 12 \pi \int_0^{M_\tau^2} {ds \over M_\tau^2}
	\left( 1 - {s \over M_\tau^2} \right)
\Bigg( {\rm Im} \Pi^{(1)}(s + i \epsilon)
	\left(  1 + P \:+\: 2 {s \over M_\tau^2} (1 - P) \right)
\nonumber \\
\;+\; {\rm Im} \Pi^{(0)} (s + i \epsilon) \left( 1 \;+\; P \right) \Bigg) \;.
\label{Dint} \end{eqnarray}
We define the energy asymmetry function $A(z)$ by
\begin{equation}{
D_P(z) \;=\; D_0(z) \; ( 1 \;+\; A(z) P ) \;,
} \label{DP} \end{equation}
where $D_0(z)$ is the energy distribution for an unpolarized tau.

By the same arguments that were used for $R_\tau$ and $A_\tau$,
the quantities $D_0(1)$ and $A(1)$ can be calculated rigorously using QCD.
Perturbative corrections can be expanded as a power series in
$\alpha_s(M_\tau)$, and nonperturbative corrections can be expanded
systematically in powers of $1/M_\tau^2$.
The free quark prediction are $D_0(1) = 5$ and $A(1) = 1/5$.
We express the QCD corrections in terms of fractional corrections
$\delta^{(n)}_R$ and $\delta^{(n)}_L$ which scale like $1/M_\tau^n$:
\begin{equation} {
D_{P=+1}(z=1) \;=\;  6 \; S_{EW} \;
\left(1 \;+\; \delta^{(0)}_R \;+\; \delta^{(2)}_R \;+\; \delta^{(4)}_R
	\;+\; \delta^{(6)}_R \;+\; ... \right) \;,
} \label{DRdel} \end{equation}
\begin{equation} {
D_{P=-1}(z=1) \;=\;  4 \; S_{EW} \;
\left(1 \;+\; \delta^{(0)}_L \;+\; \delta^{(2)}_L \;+\; \delta^{(4)}_L
	\;+\; \delta^{(6)}_L \;+\; ... \right) \;.
} \label{DLdel} \end{equation}
The fractional corrections due to the perturbative interactions
of massless quarks and gluons are
\begin{eqnarray}
\delta^{(0)}_R &=&  {\alpha_s \over \pi}
\;+\; 5.015 \left( {\alpha_s \over \pi} \right)^2
\;+\; 24.50 \left( {\alpha_s \over \pi} \right)^3
\;+\; (d_4 + 68.7) \left( {\alpha_s \over \pi} \right)^4 \;,
\label{delR0} \\
\delta^{(0)}_L &=&  {\alpha_s \over \pi}
\;+\; 3.515 \left( {\alpha_s \over \pi} \right)^2
\;+\; 4.54 \left( {\alpha_s \over \pi} \right)^3
\;+\; (d_4 - 123.9) \left( {\alpha_s \over \pi} \right)^4 \;.
\label{delL0} \end{eqnarray}
The fractional corrections from the running strange quark mass are
\begin{eqnarray}
\delta^{(2)}_R &=&  - 6 \; \sin^2 \theta_C
	\; \left( 1 + {13 \over 3} {\alpha_s \over \pi} \right)
	\; {m_s^2 \over M_\tau^2} \;,
\label{delR2} \\
\delta^{(2)}_L &=&  - {9 \over 2} \; \sin^2 \theta_C
	\; \left( 1 + {14 \over 3} {\alpha_s \over \pi} \right)
	\; {m_s^2 \over M_\tau^2} \;.
\label{delL2} \end{eqnarray}
At the order in $\alpha_s(M_\tau)$ to which we have calculated the
coefficients of the matrix elements,
the fractional corrections $\delta^{(n)}_R$ and $\delta^{(n)}_L$,
$n=4,6,$ are related in a simple way to the fractional
corrections to $R_F$ and $R_B$.  The dimension-4 corrections are
$\delta^{(4)}_R = \delta^{(4)}_F/3$ and $\delta^{(4)}_L = \delta^{(4)}_B/2$
and the dimension-6 corrections are $\delta^{(6)}_R = 0$ and
$\delta^{(6)}_L = \delta^{(6)}_B/4$.
The QCD predictions for $D_0(1)$ and $A(1)$ as a function of
$R_\tau$ are presented in Table 1.
The error on $D_0(1)$ is remarkably small, only $0.4 \%$ for
$R_\tau = 3.60$.  The dominant errors are $0.3 \%$ from
$<\!\!(\alpha_s/\pi) GG\!\!>$, $0.3 \%$ from $d_4$,
and $0.1 \%$ from $\rho \alpha_s\!<\!\!{\bar \psi}\psi\!\!>^2$.
The error in $A(1)$ is dominated by the gluon condensate and is
$ 4.5 \%$ for $R_\tau = 3.60$.  The next largest errors are $1.1 \%$
{}from $d_4$ and $0.9 \%$ from $\rho \alpha_s\!<\!\!\bar \psi \psi\!\!>^2$.
Taking $R_\tau = 3.607 \pm 0.036$ and combining the error from Table 1
in quadrature with the error due to $R_\tau$, the QCD predictions are
$D_0(1) = 5.853 \pm 0.051$ and $A(1) = 0.246 \pm 0.011$.

The QCD predictions for $A_\tau$ and for the high energy parameters
$D_0(1)$ and $A(1)$
are sufficiently precise that measurements of these parameters could be used
to determine the polarization of a sample of taus.  The polarization of
the taus produced in the decay of the $Z^0$ has been measured
using the energy distributions of the
exclusive decay modes $e^- {\bar\nu}_e \nu_\tau$,
$\mu^- {\bar\nu}_\mu \nu_\tau$, $\pi^- \nu_\tau$, $\rho^- \nu_\tau$,
and $a_1^- \nu_\tau$ \cite{lep}.
The measurement of the endpoint value $D_0(1)(1 + A(1) P)$
of the inclusive hadronic energy distribution
might be competitive as a method for determining
the tau polarization $P$, since it is
less sensitive to errors due to particle identification.

Precise measurements of the asymmetry parameters could also be used to
determine the nonperturbative matrix elements that arise in the operator
product expansion.  The errors in $A_\tau$ and $A(1)$ are dominated by
the gluon condensate $<\!\!(\alpha_s/\pi)GG\!\!>$,
with the other errors being smaller by a factor of 3.
Thus precision measurements of $A_\tau$ and $A(1)$ could
improve the determination of the gluon condensate by about a factor of 3.

In this Letter, we have introduced several new observables involving
hadronic decays of the tau lepton that can be calculated rigorously from QCD.
The only ones that were known previously were the ratio $R_\tau$ and the
moments of the invariant mass distribution $dR_\tau/ds$ \cite{eba,ldp}.
The endpoint value of the hadronic energy distribution $dR_\tau/dz$ in the
decay
of a high energy tau can also be calculated rigorously, as can the moments
of the energy distribution \cite{ebc}.  For each of these quantities,
there is also a polarization asymmetry that can be calculated.
The QCD predictions for the quantities $A_\tau$, $D_0(1)$, and $A(1)$
differ from the free quark predictions by about $20 \%$, which is much
larger than the errors in the QCD predictions.  Thus hadronic decays of
the tau lepton provide a remarkable laboratory in which
QCD can be tested at low energies with unprecedented precision.

This work was supported in part by the U.S. Department of Energy, Division
of High Energy Physics, under Grant DE-FG02-91-ER40684.

\vfill \eject

\bigskip
\noindent{\Large \bf Table Captions}

\begin{enumerate}
\item QCD predictions for $\alpha_s(M_\tau)$, $A_\tau$, $D_0(1)$, and $A(1)$
as a function of the ratio $R_\tau$.  The errors due to variations
of $d_4$, $m_s(M_\tau)$, $<\!\!(\alpha_s/\pi)GG\!\!>$,
$<\!\!m{\bar \psi}\psi\!\!>$, and $\rho \alpha_s\!<\!\!{\bar \psi}\psi\!\!>^2$
have been added in quadrature.
\end{enumerate}

\vskip 2in

\renewcommand{\arraystretch}{1.5}
\centerline{
\begin{tabular}{|c|c|c|c|c|}
\hline
$R_\tau$ & $\alpha_s(M_\tau)$ & $A_\tau$ & $D_0(1)$ & $A(1)$ \\
\hline
3.50  &$ 0.287 \pm 0.009 $&$ 0.407 \pm 0.023 $&
		$ 5.735 \pm 0.022 $&$ 0.241 \pm 0.011 $\\
3.52  &$ 0.294 \pm 0.009 $&$ 0.408 \pm 0.023 $&
		$ 5.761 \pm 0.022 $&$ 0.242 \pm 0.011 $\\
3.54  &$ 0.301 \pm 0.010 $&$ 0.409 \pm 0.023 $&
		$ 5.786 \pm 0.023 $&$ 0.243 \pm 0.011 $\\
3.56  &$ 0.308 \pm 0.010 $&$ 0.411 \pm 0.023 $&
		$ 5.812 \pm 0.023 $&$ 0.245 \pm 0.011 $\\
3.58  &$ 0.315 \pm 0.011 $&$ 0.412 \pm 0.023 $&
		$ 5.837 \pm 0.024 $&$ 0.246 \pm 0.011 $\\
3.60  &$ 0.321 \pm 0.012 $&$ 0.413 \pm 0.022 $&
		$ 5.862 \pm 0.025 $&$ 0.247 \pm 0.011 $\\
3.62  &$ 0.328 \pm 0.012 $&$ 0.414 \pm 0.022 $&
		$ 5.886 \pm 0.026 $&$ 0.248 \pm 0.011 $\\
3.64  &$ 0.334 \pm 0.013 $&$ 0.415 \pm 0.022 $&
		$ 5.911 \pm 0.027 $&$ 0.250 \pm 0.011 $\\
3.66  &$ 0.340 \pm 0.013 $&$ 0.417 \pm 0.022 $&
		$ 5.935 \pm 0.028 $&$ 0.251 \pm 0.011 $\\
3.68  &$ 0.346 \pm 0.014 $&$ 0.418 \pm 0.022 $&
		$ 5.959 \pm 0.029 $&$ 0.252 \pm 0.011 $\\
3.70  &$ 0.352 \pm 0.015 $&$ 0.419 \pm 0.022 $&
		$ 5.983 \pm 0.031 $&$ 0.254 \pm 0.011 $\\
\hline
\end{tabular}}

 \vskip 0.7 true cm
 \vbox{ \baselineskip=6 mm
        \noindent
        \centerline{Table 1} }

 \vskip 1 cm
\bigskip

\end{document}